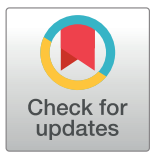

RESEARCH ARTICLE

# Perceived university support and environment as a factor of entrepreneurial intention: Evidence from Western Transdanubia Region

Attila Lajos Makai[1]*, Tibor Dőry[2]

1 Doctoral School of Regional- and Business Administration Sciences, Széchenyi István University, Győr, Hungary, 2 Kautz Gyula Faculty of Business and Economics, Széchenyi István University, Győr, Hungary

* makai.attila.lajos@sze.hu

## Abstract

The exploration of entrepreneurship has become a priority for scientific research in recent years. Understanding this phenomenon is particularly important for the transformation of entrepreneurship into action, which is a key factor in early-stage entrepreneurial activity. This gains particular relevance in the university environment, where, in addition to the conventional teaching and research functions, the entrepreneurial university operation based on open innovation, as well as the enhancement of entrepreneurial attitudes of researchers and students, are receiving increased attention. This study is based on a survey conducted among students attending a Hungarian university of applied science in Western Transdanubia Region who have demonstrated their existing entrepreneurial commitment by joining a national startup training and incubation programme. The main research question of the study is to what extent student entrepreneurship intention is influenced by the environment of the entrepreneurial university ecosystem and the support services available at the university. A further question is whether these factors are able to mitigate the negative effects of internal cognitive and external barriers by enhancing entrepreneurial attitudes and perceived behavioural control. The relatively large number of students involved in the programme allows the data to be analysed using SEM modelling. The results indicate a strong covariance between the perceived university support and environment among students. Another observation is the distinct effect of these institutional factors on perceived behavioural control of students.

## 1. Introduction

Recognizing the role of entrepreneurship in exploring new opportunities is not a new scientific discovery. This was first introduced into economic sciences by the advocates of the Austrian school of economics [1]. The phenomenon that Hayek identifies as "knowledge problem" is to be found in the coordination of different complementary individual skills rather than in the allocation of different resources, consumer preferences or technologies [2]. The key drivers of these market mechanisms are entrepreneurs who examine consumers, resources, and competitors in a constant "entrepreneurial process" in order to discover new profitable opportunities. In practice, entrepreneurs are a group of society that can effectively adapt to unforeseen changes within the economic system, thereby ensuring a continuous value creation for the community [3]. The importance of entrepreneurial mindset and related skills is not limited to conventional entrepreneurial activity: their role in intrapreneurship-based innovative development within large enterprises is inevitable [4], and, furthermore, they also form a basis for intervention in the implementation of certain policies [5].

Entrepreneurial drive and commitment are certainly not isolated and context-free phenomena. Similar to other factors, they are strongly influenced and shaped by environmental conditions. It is no coincidence that an ecosystem approach to the analysis of entrepreneurship and innovation systems [6, 7] is gaining ground in the literature. Higher education institutions are key players in entrepreneurial ecosystems [8] contributing to the enhancement of entrepreneurial drive through academic entrepreneurship and entrepreneurial education functions [9]. The ecosystem and the context-based approach also play a central role in research related to the exploration of the components of entrepreneurial intentions. Based on their results, Tolentino et al. draw attention to this aspect, highlighting the significant impact of the contextual and situational effects on entrepreneurial intention [10]. Systematic literature reviews analysing the academic discussion of the topic also underline the importance of context [11–14].

For the analysis of this topic, many different and very diverse models have been developed combining the contextual elements into a single system. Of these, the conceptual framework for the present study is based on the Theory of Planned Behaviour (TPB), developed [15] and later enhanced by Ajzen [16]. Based on this, a substantial amount of studies have been carried out in the past, confirming the effectiveness of the model in predicting entrepreneurial intentions [17–22]. By assigning each element of the context to specific variables (such as: constraints, external environment, university support), the Entrepreneurial Intention-Constraint Model (EICM), a well-practicable conceptual model created by Trivedi [23], provides the framework for this research. The validity of the original framework was demonstrated in the referenced study, where the data were collected from management students in Malaysia, India and Singapore. The survey that provides the basis for this study was conducted among those students at Széchenyi István University in Hungary who previously applied to the Hungarian Startup University Program (HSUP) organised and funded by the state. The programme itself is basically a multifunctional initiative to strengthen startup competences and knowledge. The course is fully online, and the curriculum was developed centrally by the Hungarian Express Innovation Agency. However, the coordination tasks have been localized at university level and not centrally. The specificity of the two-semester programme is that it goes beyond the conventional entrepreneurial education framework, as from the second semester it focuses on the development and mentoring of teams formed by students on a voluntary basis. During the semester, these teams create a PoC (Proof of Concept) for a project they have been developing, as well as a marketable business model and business plan. Their work is supported by technical and financial advisers and mentors who help them achieve the best possible results. At the end of the programme, the successful teams will have the opportunity to receive incubation venture investment or angel investment and will also be assisted in pitching for this within the framework of the programme. Due to the characteristics of the programme, students who are planning to set up a startup or those who already have a concrete development plan are typically enrolled, so the entrepreneurial drive among them is relatively strong. It should be noted here that not all participants of the survey responded that they plan to start a business after the course. This makes a difference compared to university accelerator or incubation programs, where the main goal of each participant is typically to launch a business. The HSUP program could be considered as a hybrid initiative where entrepreneurial education and business incubation functions are combined.

The research presented in the article focuses on university students with relatively strong entrepreneurial intentions and examines the effects of the perceived university environment and the support services of the university on entrepreneurial intentions and entrepreneurial attitudes, considering other contextual elements. Previous research results [24–26] have proved the effect of entrepreneurial education and mentoring activities on the stimulation of becoming an entrepreneur. At the same time, the mere education function on entrepreneurship does not result in an increase in the number of businesses started by university students. The process is much more complex where the influence of higher education environment cannot be neglected. Systemic literature reviews on students' entrepreneurial intentions stress the importance of several factors and highlight the quality of university environment and support system available for students. In addition, scholars suggest to investigate how combined factors such as environmental factors, contextual factors, and social factors affect entrepreneurial intentions of university students [27, 28].

The aim of the paper is twofold. First, is to provide insights about the influencing mechanisms of higher education on student entrepreneurial intention in a specific socio-economic and geographic context (Western Transdanubia Region, Hungary.) Second, is to gain a better understanding of student's perceptions on campus environment which is highly influenced with the effects of entrepreneurial model shift process nowadays in Hungary. In our view, the survey results in an emerging innovator country [29] could also be relevant for stakeholders of the innovation ecosystem in other countries and regions, beyond the European Union.

This study is structured as follows: First, it identifies what factors influence entrepreneurial drive and discusses the relevant literature on entrepreneurial intention and formulates the main hypotheses to be tested. The next section presents the research method i.e., the applied SEM method, the variables and their structuring. In the last section, the results are presented along with an introduction to testing the hypotheses. Due to the subject-matter of the research, the final sections present a range of possible limitations, as the results may not be applied to the entire population of university students.

## 2. Review of the literature

Entrepreneurial intention is an area that has been analysed and has been debated on many aspects for decades in entrepreneurship literature. This is a rapidly evolving and expanding research topic based on the findings of related systematic literature reviews [11, 12, 30]. In the light of cited literature analyses, there is a consensus can be observed that becoming an entrepreneur is a complex process where behaviour, motivation, other subjective factors, and external environmental elements have important role. These factors form virtually infinite combination, causing difficulties to identify the most important ones. The popularity of conceptual models supporting the structuring process can be explained by the mentioned complex nature of the phenomenon. The theoretical foundations of models related to the explanation of entrepreneurial activity are determined by three major frameworks. The Social Cognitive Theory emphasises the effects of entrepreneurial traits, behaviour and the external environment on entrepreneurial activity and intentions [31]. The Entrepreneurial Event model emphasizes the importance of the effects on the person starting the business [32]. These effects can be positive or negative, but in the absence of them, the entrepreneurial turn most likely will not occur in a person's life. At the same time, the entrepreneurial event is influenced by several other factors (like how desirable the person finds being an entrepreneur). In addition, attitudes are indirectly influenced by factors such as previous entrepreneurial experience, previous job and role models. Another popular model is Ajzen's Theory of Planned Behaviour [15]. According to the model, becoming an entrepreneur is explained by the intention to become an entrepreneur. Entrepreneurial intention is a function of three factors, intention, attitude, and perceived behaviour control. Since the present research focuses on entrepreneurial intention, the applied theoretical framework is based on Ajzen's theory.

Intention is a factor that affects the realization of entrepreneurial action in the model. The progress from intention to behaviour requires entrepreneurs to disentangle many inconveniences linked to implementation. Individuals may delay behaviour or do not act due to some factors. Therefore, the importance of understanding the influence of entrepreneurial intention on behaviour is crucial. Several additions to the model have been published in the literature in recent years. While searching for greater explanation several variables have been introduced, such as anticipated ambivalence [33], social valuation [34], endogenous and exogenous barriers [23], entrepreneurial knowledge [35] and creativity [36].

Research on the impact of the university as a specific entrepreneurial ecosystem [37] also appears in the literature in several aspects. Morris et al. [38] examined the impact of the university ecosystem on influencing entrepreneurial intention of students. Based on their results, extra-curricular entrepreneurship education had a positive effect on becoming an entrepreneur, while the financial support received from the university had a negative effect. Other research related to the topic found no evidence of a significant effect of the university environment on entrepreneurial intention [39]. Passoni and Glavam [40] verified in a population of Brazilian students majoring in management, engineering and accounting that entrepreneurship education had a positive effect on the willingness of management and engineering students to start a business. In case of student start-up activities, Bergmann and his co-authors [26] made an attempt to measure the influence of factors affecting entrepreneurial intention at the individual, university and regional level. According to their findings, nascent enterprises are more influenced by the university environment and the influence of the regional level was insignificant, the opposite can be seen in the case of already active student enterprises. Based on a Brazilian sample, Canever et al. [41] verified that no significant differences can be detected in the entrepreneurship willingness of students of public and private universities. Silva et al. [42] proved that university entrepreneurship education has a stimulating effect on students becoming entrepreneurs, and the university atmosphere and entrepreneurial skills also play an important role. However, they did not examine the role of university support in their research. Dvorský et al. [43] discovered regional differences in the Central and Eastern European context regarding the impact of universities on entrepreneurial intention. It is also a Central Eastern European conclusion that the direct impact of entrepreneurship education on intention can be demonstrated to a significant degree in countries where entrepreneurship education is taking place at the secondary school level. As can be seen from the (non-systematic) literature review, the complex area of entrepreneurial intention and the factors influencing it has been examined from many perspectives. It can be clearly stated that the influence of the higher education institutional environment on intention (and the method of this) is a subject of debate in the literature. It is also evident that the geographical context has significant importance.

## 3. Theoretical framework

The key element of Ajzen's Theory of Planned Behaviour (TPB) [15] is that it distinguishes between intention and action. However, the existence of entrepreneurial intention is only a necessary but not sufficient condition for entrepreneurial action, i.e., (launching) setting up a business. According to the model, the degree of entrepreneurial intention itself is dependent on the combined effect of multiple other components. In this respect, entrepreneurial attitudes, perceived social standards and observed behavioural control can be highlighted. At the same time, entrepreneurial intention and action are also influenced by objective factors, which in the later version of this model [16] are called actual controls. There are various factors at disposal in a particular entrepreneurial ecosystem, such as the current availability of the necessary resources, competences, and other support. There is a positive correlation between entrepreneurial attitude toward behaviour and entrepreneurial intention in the model. This relation has been confirmed by a number of research and studies [44–47]. This connection must necessarily be strong and decisive also in the target group of the present research, where the application to the programme can already be interpreted as a kind of "entrepreneurial action", since (as highlighted earlier) the HSUP programme, due to its incubation elements, goes beyond a conventional programme that aims at enhancing entrepreneurial skills. For these reasons, Hypothesis 1 contains the following statement:

H1: The entrepreneurial intentions (EI) of the surveyed students are (very) strongly influenced by their existing attitude toward behaviour (ATB).

Within the framework of the Ajzen model, particular attention is paid to perceived social/subjective norms denoting a set of family, kinship and friend opinions related to the launch of a business and serving as a reference. In this context, there have been rather controversial research findings. Kolvereid and Isaksen identified a weak correlation between entrepreneurial attitude and perceived social standards [48]. Franke and Lüthje also found a weak relation between the variables [44]. In contrast, several other studies did not find a significant relationship between the two variables [19, 46, 49, 50]. This suggests that there may be an indirect (and presumably weak) relation between the two variables. The associated hypothesis can be summarized as follows:

H2: Social norms perceived by the surveyed students (PSN) positively influence their attitude toward behaviour (ATB) and perceived behavioural control (PBC).

The impact of the perceived behavioural control (PBC) on the business startup intention has already been demonstrated in relation to the original Ajzen model, but this correlation is supported also by several research [19, 49–53] findings. It is noteworthy that the original Ajzen model includes covariances among the factors influencing the intention to start a business (attitude, perceived subjective norms, perceived behavioural control); however, based on research it is questionable whether these explanatory variables are independent of each other, which makes it necessary to treat the interacting covariances with reservations [50]. In view of the above, the related hypothesis is defined as follows:

H3: Perceived behavioural control (PBC) positively affects both entrepreneurial intention (EI) and attitude toward behaviour (ATB).

Trivedi points out [23] that there is a paucity of empirical research on the role of specific constraints on entrepreneurial commitment and intention among startup entrepreneurs. These may be external and internal (cognitive) factors, depending on whether they are outside or inside the control of the subject [54]. External constraints can be economic, political or regional by nature, and in contextual terms, all have a perceived impact on entrepreneurial intentions [55], in relation to which empirical studies were conducted regarding the impact of several elements [44]. The potential impact of external barriers on internal constraints, which deserves further investigation. Accordingly, the related hypothesis of this study includes two elements:

H4/A: Internal (cognitive) barriers (COG_BARR) have a negative impact on attitude toward behaviour (ATB) and perceived behavioural controls (PBC).

H4/B: External barriers (EXT_BARR) strengthen and positively influence internal barriers (COG_BARR).

The role of the university environment and the entrepreneurial-university ecosystem are shaping entrepreneurial attitudes and commitment constitutes a particularly well-researched area. There are several theoretical models related to the role of higher education institutions in the ecosystem [56, 57], which basically distinguish between two very distinct elements with respect to the entrepreneurial mindset.

The campus environment includes curricular and extra-curricular training activities of universities to enhance entrepreneurial skills, as well as events and programmes related to entrepreneurial activities, and, furthermore, the attitude, culture and atmosphere of the higher education institution towards entrepreneurship and entrepreneurial lifestyle [58–61]. The positive impact of this function on entrepreneurial attitudes and intentions has also been widely demonstrated in research studies [51, 62–66].

University support goes substantially beyond the campus atmosphere and educational activities, as it includes specific forms of support provided by the university to businesses (advisory services, provision of capital investment, support for market entry) [67]. In this respect, the university acts as a partner and stakeholder and the success of the supported business represents an economic interest also for the university [68–75]. Within the framework of the original EICM model, [23], the university environment was included as a variable in the model, but in view of the above, the present research assumes the following hypotheses regarding the effects brought about by the university:

H5/A: A supportive university environment (UNI_ENV) has a positive impact on students' attitude towards becoming entrepreneurs (ATB).

H5/B: The university's business support services (UNI_SUPP) have a positive impact on students' perceived behavioural control (PBC).

As explained above, the theoretical framework of the present research is a slightly modified EICM model. This provides a suitable tool for the purposes of the research, i.e. for examining the effects of the university environment and support. However, it should be noted that the applied model has several weaknesses, which appear as a limiting factor in the interpretation of the results. One of the most important weaknesses of the model is that it does not include variables related to creativity, anticipated ambivalence and social valuation among the factors that affect entrepreneurial intention. Another weakness is that it does not analyse the relationship between entrepreneurial action and intention, which is also influenced by many factors.

The hypotheses identified by the research are shown in (Fig 1):

## 4. Research method

The survey underlying this research took place in November and December 2021. The population consisted of students studying at Széchenyi István University of Győr, Hungary who had enrolled in the aforementioned HSUP programme. Széchenyi István University is the largest higher education institution in North Transdanubia. This region has the most developed agricultural and food industry in the country, but it is clearly the industrial sector that plays a leading role and has fostered dynamic growth and development since the second half of the 1990s. The northern part of the region, mainly Győr-Moson-Sopron county and to a small extent Vas county, is geographically well integrated into the European and world market, and its comparative advantage still lies in its industry (mainly machinery and automotive). The latest important step in the transformation of the entrepreneurial university is the process known as the higher education model change, which aims to create an efficient and modern higher education system in Hungary. In the process, the founding and maintenance rights of the model-

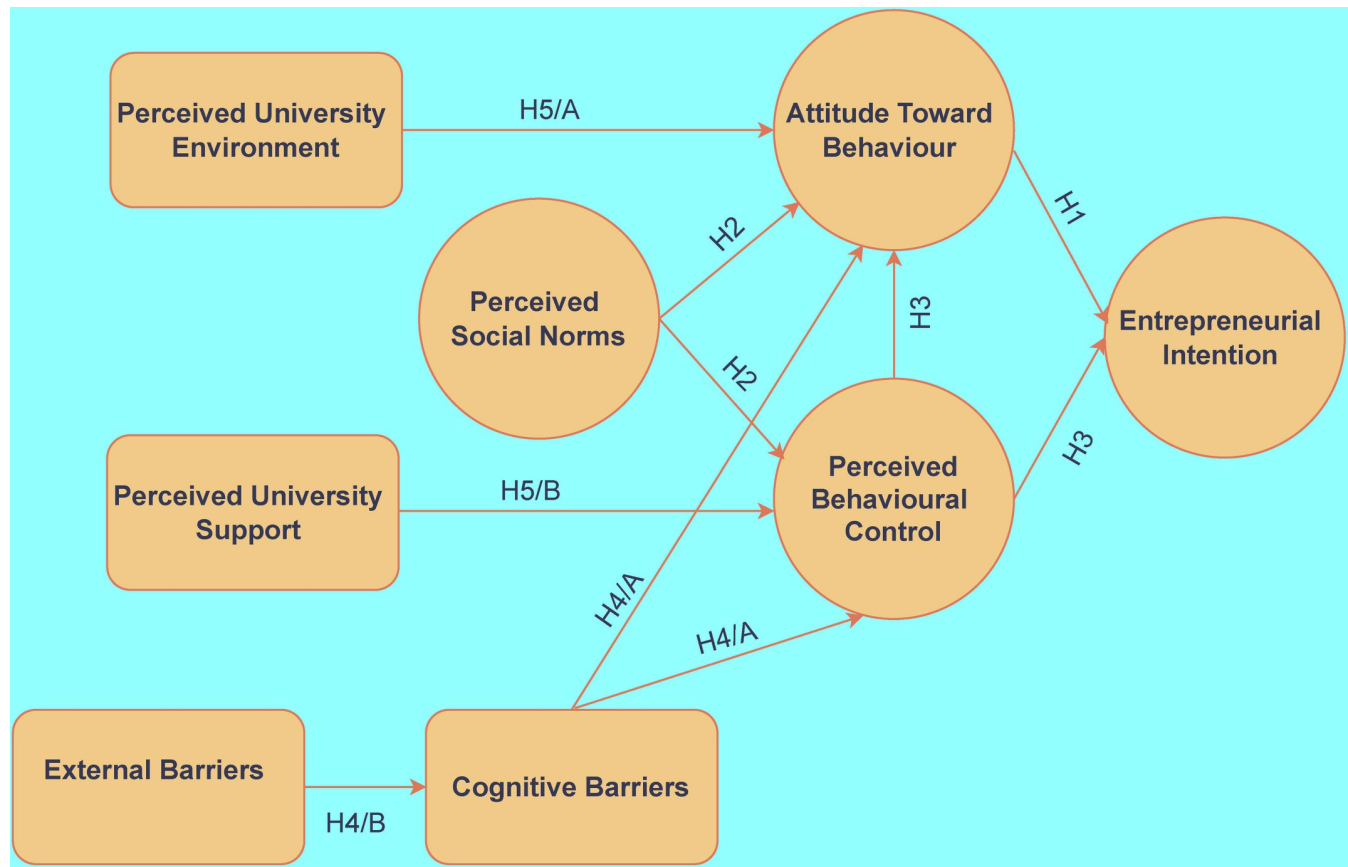

**Fig 1. Conceptual framework & hypotheses.**

https://doi.org/10.1371/journal.pone.0283850.g001

changing institutions were transferred from direct state ownership to trust foundations, which created new opportunities for technology-focused universities such as Széchenyi István University.

In the academic year 2021–2022, 480 students from Széchenyi István University applied for the HSUP programme. This group of students was sent an anonymous questionnaire with full respect of the data protection rules in force by means of GoogleForms. A total of 145 respondents (N = 145) returned the questionnaire within the 30 days given to complete the survey, which is a fairly good response rate compared to the baseline population. Based on the relevant methodological literature, the sample size should be greater than 10 times the maximum number of inner- and outer model links pointing at any latent variable in the model [76, 77]. At the same time, when SEM models are being applied to relatively small sample size, reducing the acceptability of the significance level to ≤0.10 may be considered reasonable [76]. The current number of respondents exceeds the minimum number of elements required for SEM modelling in relation to the links pointing at used variables. The received answers were checked manually. This process was also helped by the function of GoogleForms, which does not allow the respondent to move on to the next question in the case of missing answers. Therefore, the received surveys did not contain missing variables. Respondents included 83 men (57.24%) and 62 women (42.76%). 73 people (50.33%) were from the technical education area and 72 people (49.67%) from the business administration area. A significant proportion of students (116 students, 80%) were enrolled in state-funded programmes, while a smaller rate (29

Table 1. Demographic information about respondents.

| Variable | | Frequency | Percent | Std. Deviation | Variance |
|---|---|---|---|---|---|
| Gender | Male | 83 | 57.2 | .496 | .246 |
| | Female | 62 | 42.8 | | |
| Age | 1970–1979 | 2 | 1.4 | .725 | .526 |
| | 1995–1999 | 71 | 49 | | |
| | 2000–2004 | 72 | 49.7 | | |
| Residence | Budapest | 7 | 4.8 | 1.346 | 1.811 |
| | Győr (county center) | 34 | 23.4 | | |
| | Other county center | 16 | 11 | | |
| | Town | 37 | 25.5 | | |
| | Village | 47 | 32.4 | | |
| | Foreign town | 4 | 2.8 | | |
| Major | Technical sciences | 73 | 50.3 | .502 | .252 |
| | Economic sciences | 72 | 49.7 | | |
| Finance | State-funded | 116 | 80 | .401 | .161 |
| | Self-funded | 29 | 20 | | |
| Dual training | Yes | 3 | 2.1 | .143 | .020 |
| | No | 142 | 97.9 | | |
| Expected graduation | 2022 | 46 | 31.7 | .757 | .543 |
| | 2023 | 62 | 42.8 | | |
| | 2024 | 37 | 25.5 | | |

https://doi.org/10.1371/journal.pone.0283850.t001

students, 20%) studied at the university on a fee-paying basis. Among the respondents, 72 students (49.67%) were aged between 18 and 22 and 71 students (49.96%) were aged between 23 and 27 (2 respondents gave unrealistic ages). Table 1 summarises the main demographics of the participant students.

The conducted research can basically be divided into 4 major sections. The demographic characteristics of the respondents (Section 1). This section contains the control variables of the questionnaire. The research basically operated with 3 control variables (age, specialization and gender). Age and specialization did not significantly influence the results and the model, but it can be observed that the entrepreneurial intention of female university students is slightly lower compared to male students. Observing and analysing gender-based differences were not the focus of current research, however the results confirm the insights of previous articles and systematic literature reviews on this topic [11, 12, 30, 78] The following questions related to the independent variables (Section 2: entrepreneurial intention [EI], entrepreneurial attitude toward behaviour [ATB] and perceived behavioural control [PBC] variables), then the independent or partially independent variables (Section 3: cognitive barriers [COG_BARR], external barriers [EXT_BARR], perceived social norms [PSN], university environment and support variables [UNI_ENV and UNI_SUPP]). In Section 4, other questions are included that are not in connection with the present research but relate to the future use of the university services and are for internal use.

The set of questions selected for the independent variables is mainly based on questions used and applied in previous surveys with similar objectives [52], as well as on the original questions used to test the EICM model [23]. For the external barriers, the questions previously used for this purpose were applied [44], and for the internal barriers, the research questions related to the EICM model were adopted [23]. The perceived social norms were also measured through questions that were used previously for similar purposes. [50] For the university environment and support, the questions developed by Kraaijenbrink et al. [79] to measure the perceived campus environment and support among students at European universities were used. For each question, a five-point Likert scale was provided for response (in the range from "Strongly agree" to "Strongly disagree").

In order to evaluate the responses from the completed questionnaires, following the data transformation procedures required to run the IBM-SPSS analysis, we first applied factor analysis with varimax rotation and Kaiser normalisation to extract the specific variables.

To check for adequate reliability, the item-to-total correlation should be at least 0.35 for questions related to a single variable, and the Cronbach's Alpha value should exceed 0.70, as recommended in the related literature [80, 81]. It is also necessary to analyse the factorability of the set of variables. The 0.861 value of the KMO test (Kaiser-Meyer-Olkin index) for the whole sample is considered very good (KMO > 8). The Bartlet spherical test also shows significance. The 72.352% value of total variance explained (TVE) is acceptable [81]. Table 2 summarises the results of the reliability analysis.

In terms of validity, both convergent and discriminant validity were checked on the basis of literature recommendations [81]. For composite reliability (CR) and average expressed variance (AVE), the published threshold values (CR > 0.7; CR > AVE; AVE > 0.5) have been used as a reference for [81, 82] to check convergent validity. It must be noted that the range of the factor weights associated with each variable also exceeds 0.5 in all cases. Table 3 contains the indices related to convergent validity.

Certainly, the literature also provides guidance on discriminant validity requirements [82]. These can be summarised as follows:

a. The root of the expressed average variance of a given factor exceeds the correlation coefficient between the respective factor and all other factors.

b. The maximum split variance (MSV) value shall be less than the average variance expressed (AVE) by the factors. That is: MSV < AVE.

c. The value of the average shared variance (ASV) shall be less than that of the variance expressed by the factors (AVE). That is: ASV < AVE.

The indices of discriminant validity are summarised in Table 4.

Overall, it can be concluded that the reliability and validity analyses confirm the hypotheses that the questions used and the variables derived are suitable for the assessment of correlations. Following the goodness-of-fit analysis, the results and the model outlined using the SEM method are discussed below.

## 5. Analyses and findings

The primary objective of the research is to examine the relation between entrepreneurial intention (as a dependent variable) and the predictors defined in the EICM model on a sample in which the entrepreneurial commitment of the population can be considered high. In order to be able to test the model on the basis of the variables presented in the previous section, it is first necessary to make sure that the goodness-of-fit is guaranteed. Based on the literature recommendations, the fit itself can basically be examined in three aspects [83–85]:

a. Absolute fit, where the focus of attention should be placed on the GFI (goodness-of-fit) index, the RMSR (root mean square residual) and the RMSEA (root mean square error of approximation).

b. Incremental fit, where the TLI (Tucker-Lewis index), IFI (incremental fit index) and CFI (comparative fit index) indicators are examined.

Table 2. Reliability analysis.

| Variable | Question | Factor loading | Cronbach's Alpha | KMO |
|---|---|---|---|---|
| Entrepreneurial Intention (EI) | I am ready to do anything to become an entrepreneur | .872 | .894 | 0.861 |
| | I study so that I can use the acquired knowledge in my own company | .713 | | |
| | I feel I am a "born" entrepreneur | .869 | | |
| | At present, I strongly consider starting a company | .862 | | |
| | I am determined to start a company in the future | .875 | | |
| Attitude Toward Behaviour (ATB) | To me, becoming an entrepreneur has more advantages than disadvantages | .801 | .898 | |
| | An entrepreneurial career is attractive to me | .906 | | |
| | If I have the necessary resources and see an opportunity, I will start my own company | .839 | | |
| | Becoming an entrepreneur would bring me great satisfaction | .812 | | |
| | If I had to choose between career options, I would choose entrepreneurship | .857 | | |
| Perceived Social Norms (PSN) | Support from parents to start a business | .822 | .772 | |
| | Support from relatives to start a business | .883 | | |
| | Support from friends and acquaintances to start a business | .780 | | |
| Perceived University Support (UNI_SUPP) | The University actively helps me if I want to establish a startup company; I can get the legal and financial information I need and they can even help me find an office if necessary | .888 | .893 | |
| | The University helps me find an investor to launch my startup company | .891 | | |
| | The University organises idea contests, hackathons and competitions to find and support promising student initiatives | .812 | | |
| | If I were to establish a startup business, I think the University would help me to raise the recognition of my company and find potential customers | .888 | | |
| Perceived Behaviour Control (PBC) | Establishing and running a business would be easy for me | .828 | .816 | |
| | I am ready to establish an operational business | .861 | | |
| | I would be able to control the tasks related to starting a business | .841 | | |
| | I know the legal, financial and management tasks required to start and run a business | .678** | | |
| External Barriers (EXT_BARR1)* | I think it is difficult to find a business idea that has not yet been devised by others | .830 | .801 | |
| | Entering the market with a new product/service is difficult | .867 | | |
| | It is difficult to raise capital from banks or investors for a launching (startup) business | .650** | | |
| External Barriers (EXT_BARR2)* | I think nowadays it is hard to acquire knowledge or find a good advisor on how to establish and run a startup business | .667** | | |
| | I think taxation rules make it difficult to establish new businesses | .855 | | |
| | I observe that the state does not sufficiently support the establishment of startup businesses | .802 | | |
| | I think the legislation is too complicated and not favourable to the establishment of new businesses | .858 | | |
| Internal Barriers (INT_BARR) | I fear that I will eventually fail if I establish a company | .759 | .813 | |
| | I am not motivated to establish a startup company | .752 | | |
| | I do not think I have the necessary skills to establish a successful company | .838 | | |
| | I do not feel ready to manage the stress of being an entrepreneur | .852 | | |
| Perceived University Environment (UNI_ENV) | The creative atmosphere at the University encourages me to think about and develop new ideas for starting a business | .750 | .931 | |
| | Courses available at the University develop the leadership skills that are required for entrepreneurs | .833 | | |
| | The available courses provide the knowledge required to start a new business | .741 | | |
| | I find that I am motivated and supported by the professors of the University to establish a startup business | .817 | | |
| | I see that there are events at the University where I can meet successful entrepreneurs and learn from their experience | .828 | | |
| | I think that the University, both in education and in communication, presents entrepreneurship to students as a real and desirable career option | .866 | | |
| | I think the University will help me get acquainted with startup companies where I can do my practical training or get an internship | .813 | | |
| | At the University, I can find workshops, events and professional forums where I can acquire the knowledge required to set up startup business | .771 | | |
| | I think that at the University I would have the chance to find partners with whom we could establish a joint startup business in the future | .798 | | |

*For external inhibitors, two factors should be used as a result of the factor analysis, as this is the way to obtain the appropriate TVE value for the respective variable. Due to the nature of the questions, we named the first factor external market barriers (EXT_BARR_M) and the second factor policy barriers (EXT_BARR_P) and will refer to them in such a form in the following parts.

**Although the factor loading of marked items is >.70, we keep those items in the model, because of the relative high AVE value of the related variables.

https://doi.org/10.1371/journal.pone.0283850.t002

Table 3. Convergent validity.

|  | CR | AVE |
|---|---|---|
| EI | 0.922 | 0.706 |
| ATB | 0.925 | 0.712 |
| PBC | 0.881 | 0.651 |
| PSN | 0.868 | 0.687 |
| COG_BARR | 0.865 | 0.617 |
| EXT_BARR_M | 0.875 | 0.638 |
| EXT_BARR_P | 0.828 | 0.621 |
| UNI_SUPP | 0.925 | 0.757 |
| UNI_ENV | 0.942 | 0.644 |

https://doi.org/10.1371/journal.pone.0283850.t003

c. Parsimonial fit, where the PGFI (parsimony-adjusted goodness-of-fit index), the PCFI (parsimony-adjusted comparative fit index) and the PNFI (parsimony-adjusted normed fit index) figures are to be matched to specified thresholds.

The results of the goodness-of-fit analysis carried out by means of the SPSS-AMOS 28 software and the relevant thresholds defined in the referenced literature are summarised in Table 5.

In summary, the fit of the model is very good (Chi square = 36.986, degree of freedom: 24), with a CMIN/DF value of 1.541 (threshold <2). The P-value (Probability level) of 0.044 can also be considered good, taking into account the threshold of < 0.05. The derived variables are suitable for further analysis and drawing conclusions. As mentioned earlier, the original Ajzen model's [15] extended version called EICM [23] forms the basis of this research. We slightly modified, however, the original EICM model based on the results of the research for testing it, and we extended it in some aspects (addressing university support and environment as separate variables), from which we developed the framework presented in the hypothesis formulation section. Using the SPSS-AMOS 28 software, the structure shown in Fig 2 was devised for the relations between the variables.

The analysis leads to the following conclusions:

1. Entrepreneurial intention (EI), as the basic dependent variable of the research, is highly influenced by independent and partially independent variables. The resulting $R^2 = 0.68$

Table 4. Discriminant validity*.

|  | EI | ATB | PBC | PSN | COG_BARR | EXT_BARR_M | EXT_BARR_P | UNI_SUPP | UNI_ENV |
|---|---|---|---|---|---|---|---|---|---|
| EI | (0.840) | | | | | | | | |
| ATB | 0.808 | (0.843) | | | | | | | |
| PBC | 0.620 | 0.612 | (0.806) | | | | | | |
| PSN | 0.256 | 0.256 | 0.068 | (0.828) | | | | | |
| COG_BARR | -0.493 | -0.476 | -0.418 | -0.041 | (0.785) | | | | |
| EXT_BARR_M | -0.087 | -0.030 | -0.056 | 0.170 | 0.268 | (0.798) | | | |
| EXT_BARR_P | 0.031 | -0.022 | 0.067 | 0.061 | 0.201 | 0.000 | (0.788) | | |
| UNI_SUPP | 0.261 | 0.214 | 0.290 | 0.182 | 0.081 | -0.088 | 0.019 | (0.870) | |
| UNI_ENV | 0.192 | 0.194 | 0.172 | 0.259 | 0.097 | -0.114 | 0.036 | 0.800 | (0.802) |
| **AVE** | **0.706** | **0.712** | **0.651** | **0.687** | **0.617** | **0.638** | **0.621** | **0.757** | **0.644** |
| **MSV** | **0.497** | **0.358** | **0.551** | **0.100** | **0.353** | **0.152** | **0.163** | **0.440** | **0.440** |
| **ASV** | **0.261** | **0.251** | **0.227** | **0.047** | **0.149** | **0.031** | **0.042** | **0.088** | **0.102** |

*Square roots of AVE values shown in brackets.

https://doi.org/10.1371/journal.pone.0283850.t004

Table 5. Goodness-of-fit analysis.

| Absolute fit | Value | Threshold |
|---|---|---|
| GFI | 0.950 | > 0.95 |
| SRMR | 0.058 | < 0.08 |
| RMSEA | 0.061 | < 0.06 |
| **Incremental fit** | | |
| TLI | 0.960 | > 0.95 |
| IFI | 0.974 | > 0.9 |
| CFI | 0.973 | > 0.9 |
| **Parsimonial fit** | | |
| PGFI | 0.506 | > 0.5 |
| PCFI | 0.649 | > 0.5 |
| PNFI | 0.619 | > 0.5 |

https://doi.org/10.1371/journal.pone.0283850.t005

value with respect to EI provides clear evidence of this. Entrepreneurial intention is most strongly influenced by attitude toward behaviour (β = 0.69, *p<0.001)*, which explicitly confirms the statement defined in hypothesis *H1* of the present study.

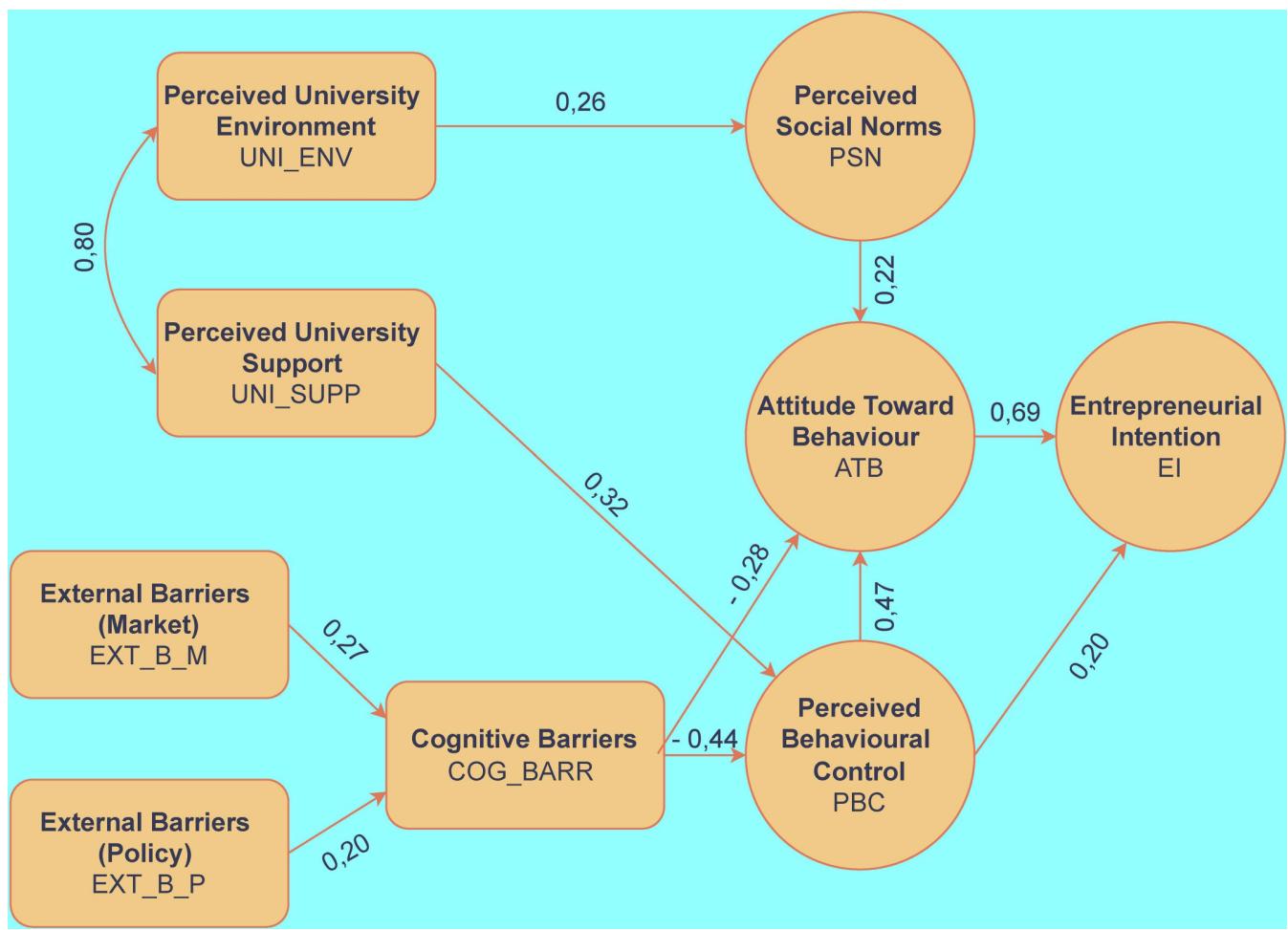

**Fig 2. Structural model result.**

https://doi.org/10.1371/journal.pone.0283850.g002

2. The social norms perceived by students (PSN) definitely influence entrepreneurial attitudes (ATB), and the effect is positive (β = 0.22, *p<0.001)*. Contrary to the results of the previous research testing the EICM model [23], no effect on perceived behavioural control (PBC) was found (β = 0.01, *p>0.05)* based on the present research results. Thus, the statements proposed in hypothesis *H2* were only *partially* confirmed.

3. Perceived behavioural control had a clearly positive effect on both entrepreneurial intention (EI) (β = 0.20, *p<0.001)* and attitude toward behaviour (ATB) (β = 0.465, *p<0.001)*. The former relation was also described in the original EICM model, the latter is a novelty based on the results of this research, which fully confirm the statements outlined in hypothesis *H3*.

4. It can be explicitly demonstrated that the cognitive (internal) barriers (COG_BARR) have negative impact (β = -0.281, *p<0.001)* both on entrepreneurial attitudes (ATB) and on perceived behavioural control (PBC) (β = -0.445, *p<0.001)*. This mechanism was also explored in the original model and confirms the statements outlined in hypothesis *H4/A*. The external barriers broken down into two factors (EXT_BARR) do not have a significant effect on attitude (ATB) or perceived behavioural control (PBC). However, based on the present research, they strengthen the internal (cognitive) barriers (COG_BARR) (Factor 1: β = 0.268, *p<0.001*, Factor 2: β = 0.201, *p<0.05)*, which are negative predictors of entrepreneurial attitude toward behaviour (ATB) and perceived behavioural control (PBC). By the indicated relation, the hypothesis *H4/B* of the present research is confirmed.

5. The university environment (UNI_ENV) has no significant effect (β = 0.084, *p>0.05)* on attitude (ATB), while perceived university support (UNI_SUPP) positively influences perceived behavioural control (PBC) (β = 0.328, *p<0.001)*. The latter is also interesting because the strength of this effect is comparable to the negative effects of internal (cognitive) barriers (COG_BARR), and as a result, the role of university support services seems to have a noticeable countervailing effect. Based on the above-mentioned fact, we can establish that the assumptions of hypothesis *H5/A* were not confirmed, whereas hypothesis *H5/B* was proven.

6. The results confirm the insights of previous research outcomes that the model fit reached the desired value only when a covariance was inserted between the variables university environment (UNI_ENV) and university support (UNI_SUPP). This particularly strong relationship (see Table 5) suggests that the examined elements of the university entrepreneurship/innovation ecosystem as perceived by the students are very closely correlated and that the perceived university support can only have a positive impact in the proper environment, which is confirmed by the verification of hypothesis *H5/B* of this research.

7. The potential impact of the university environment on perceived social norms has not been in the focus of research so far. When reviewing the relevant literature, we did not find any research of this type; therefore, no corresponding hypothesis was proposed. However, in the analyses required to develop the model (where, naturally, quite a number of variants of the model were tested), it was remarkable that, without any exception, a positive and significant relationship (β = 0.259, *p<0.001)* was found between the university environment (UNI_ENV) and the perceived social norm (PSN). Nevertheless, the resulting $R^2 = 0.07$ does not allow us to consider this as significant, yet the relationship still exists.

A summary of the regression weights and covariances associated with the model is presented in Table 6.

**Table 6. Regression weights and covariances.**

| Standardized Regression Weights | | | | | | | |
|---|---|---|---|---|---|---|---|
| | | | Estimate | S.E. | C.R. | P | Supported |
| COG_BARR | <--- | EXT_B_M | .268 | .079 | 3.407 | *** | yes |
| COG_BARR | <--- | EXT_B_P | .201 | .079 | 2.556 | .011 | yes |
| PSN | <--- | UNI_ENV | .259 | .081 | 3.212 | .001 | yes |
| PBC | <--- | COG_BARR | -.439 | .071 | -6.292 | *** | yes |
| PBC | <--- | UNI_SUPP | .322 | .072 | 4.534 | *** | yes |
| PBC | <--- | PSN | -.010 | .072 | -.134 | .894 | no |
| ATB | <--- | COG_BARR | -.263 | .066 | -4.241 | *** | yes |
| ATB | <--- | UNI_ENV | .084 | .064 | 1.324 | .185 | no |
| ATB | <--- | PBC | .489 | .068 | 6.858 | *** | yes |
| ATB | <--- | PSN | .242 | .061 | 3.623 | *** | yes |
| EI | <--- | ATB | .685 | .060 | 11.439 | *** | yes |
| EI | <--- | PBC | .202 | .060 | 3.357 | *** | yes |
| **Covariances** | | | | | | | |
| UNI_SUPP | <--> | UNI_ENV | .794 | .106 | 7.495 | *** | yes |

https://doi.org/10.1371/journal.pone.0283850.t006

## 6. Conclusion and discussion

The slightly modified EICM model, which was tested in the framework of this research, has proved to be suitable ($R^2 = 0.68$) to explain the entrepreneurial intention among university students. The research validates the essence of the original Ajzen model [15] that entrepreneurial attitude and perceived behavioural control are the factors that influence entrepreneurial intention the most. As indicated in the review of the related literature and research, the role of subjective norms in influencing entrepreneurial intentions is rather controversial. This is confirmed by the results of this research, where there is no direct correlation between these variables. A significant relationship can only be found for entrepreneurial attitudes, suggesting that perceived perceptions of the environment about starting a business can shape the entrepreneurial attitudes of university students. In this respect, the influence of relatives and family (entrepreneurial family background) is decisive, which was confirmed by former studies [52], and a systematic literature (research) review on the topic [11] also qualifies it as a dominant research direction. It is definitely a remarkable finding that the university environment has an impact (even if to a minimal extent) on the perceived social norms. This, however, comes as no surprise: apart from family and relatives, it is the opinions of friends that tend to influence the perceived social norms to a greater extent. Obviously, the network of university students' friends is concentrated on the university campus during the academic years, and a significant number of friendships are formed here. A university environment that supports and promotes entrepreneurship has an impact not only on the student who wants to start a business, but also on his or her circle of friends at university. Thus the university atmosphere can influence entrepreneurial commitment not only directly, but also through friendships that can shape the perceived social norms.

The principal novelty of the EICM model used in our research is that it includes external and internal (cognitive) barriers among the factors that shape entrepreneurial intentions and provides a useful method for measuring them. Among the internal (cognitive) factors, risk tolerance/risk-taking willingness [86] and the degree of entrepreneurial motivation play a prominent role [87, 88]. The negative influence of internal barriers is clearly confirmed by the results of the present research. A positive correlation is observed between the internal and the external inhibitors, which were not examined by the original EICM model research, as instead it focused on the direct effects of the external factors on entrepreneurial attitudes and behavioural control. The research results show that the perceived external conditions (difficulties in entering the market, insufficient government subsidies, non-transparent tax rules) increase risk aversion and reduce motivation and self-confidence among university students, which, as internal (cognitive) barriers, constitute a negative impact on entrepreneurial attitudes and perceived behavioural control.

The location, role and support of the university environment has long been a central theme in research on the factors influencing entrepreneurial intentions. The available resources and the knowledge and the services that can be accessed and directly exploited strengthen self-efficacy [89] and internal locus of control, which is an important factor for entrepreneurial and innovative attitudes [90]. By ensuring this, the innovation and entrepreneurial ecosystem of the university will significantly contribute to strengthening perceived behavioural control. This is supported by the research data and findings which show that specific university support can substantially counteract the negative effects of the internal (cognitive) barriers. Nevertheless, the results in this case do not prove that the university environment has a significant impact on entrepreneurial attitudes. The reason for this may relate to the fact that, for the purposes of the present research, the perceived university environment and the perceived university support were separated and used as two separate variables. Statistically, these could certainly be treated as a single variable, but the literature on university functions and university entrepreneurial activities largely separates these factors, as outlined in the literature review above. This is not a coincidence: internal rules, legislative constraints and management standards often prevent higher education institutions from providing capital, services and marketing support to businesses, and in many cases, it is difficult for universities to participate in the business themselves. Thus, in many cases, an entrepreneur-friendly university environment and a strong entrepreneurial education function are not accompanied by support services to foster student and researcher entrepreneurship connected to the university ecosystem. For this reason (and for more accurate research results), it is worth considering these two areas as separate variables. Yet, it is evident that the two variables move together, with a highly significant covariance between them. In other words, even if a higher education institution provides effective support services, these will only be perceived by students if these are accompanied by a corresponding perceived entrepreneur-friendly university environment. Since we are talking about covariance, the effect is naturally also true in reverse: it is useless to have an entrepreneur-friendly environment on campus if there is a lack of business support services that are present, known and perceived by students.

## 7. Limitations

Although the validity and the goodness-of-fit analyses notably show favourable values, the results of the research exhibit several limitations. The limitations are highlighted in the summaries of all the research on this topic, and it is apparent that they constitute a significant part of the explanatory power differences across the models constructed. These include the characteristics of the target groups selected for the research, national or regional specificities, temporality, and (in the case of analyses that also examine higher education-related factors) the relevant characteristics of the university/universities concerned. In view of the above considerations, the results of the present research can be interpreted along the following limitation criteria:

- The research was carried out between November and December 2021. During this period, the COVID-19 epidemic still caused significant difficulties in the region (and, of course, everywhere in the world). In 2020 and 2021, effects of the epidemic particularly affected the higher education sector, where in-person education and events became impossible to hold. However, university courses were held in person format in the Hungarian higher education system in Fall Semester 2021, including our research period. All these circumstances may have reduced the perceived negative effects of COVID in education, but it is necessary to note that events were not allowed during this period either. These conditions may have influenced the results of our research.

- The survey was conducted on a sample of students of a Hungarian, and in particular a regional university. National and regional characteristics strongly influence entrepreneurial ecosystems [91] and the related attitudes of actors towards them [92]. These effects are also noticeable on university campuses as micro-scale ecosystems [37]. With respect to Hungary (similar to the post-socialist countries), the level of entrepreneurial activity and the social recognition of entrepreneurship as an activity is lower compared to Western countries. This is also demonstrated in the university environment [93]. The moderate level of innovation activity and scarcity of resources in the country [94] have a fundamental impact on entrepreneurial ecosystems and their actors. It is recommended to interpret the obtained results in the above-mentioned context.

- The sample selected for data collection was taken from a technology-focused university. Therefore, the results obtained can only be valid for this group, while the specific characteristics of science universities in the classical sense and universities of arts are likely to have an impact on the support and university atmosphere perceived by students, thus separate research may be required for these types of institutions. A similarly influential factor is the so-called higher education model change process currently in progress in Hungary, which aims to transform the vast majority of the Hungarian higher education system into an entrepreneurial university enhancing the third mission [95].

- The model applied for the analysis has a certain number of weaknesses that limit the interpretation of the results. As discussed above, a main weakness of the model is that the factors influencing entrepreneurial intention do not incorporate variables related to creativity, expected ambivalence and social evaluation. A further weakness is that it does not analyse the relationship between entrepreneurial action and intention, which is also influenced by a number of factors that are referred to in the literature as "actual control" [16]. The essence of this concept is on the distinction between entrepreneurial intention and entrepreneurial action. In other words, even strong entrepreneurship does not necessarily result in the entrepreneurial action being taken. In the present research, the factors influencing entrepreneurial intention (attitudes, subjective norms, perceived behavioural control, perceived university support, perceived external barriers, internal barriers) were examined. At the same time, the conversion of entrepreneurial intention into action is influenced by several factors, which Ajzen called *actual control*, and which include objective factors (access to resources, sufficient time to start a business, access to market opportunities). The impact of these on the action is not addressed in the research, yet it is necessary to indicate that among the variables that belong to the actual control category (and influence action) there are several ones that overlap with the elements that influence entrepreneurial intention (external barriers, university support). The influence of these variables on specific actions was analysed on a Hungarian sample by Gubik et al. [96] and it was concluded that it contributed to the model's explanation of entrepreneurial action, but with relatively little strength. It is important to note that the referenced research was conducted before the Hungarian university model change process, therefore it does not discuss the potential impact of this process on actual behavioural control and entrepreneurial action.

## 8. Implications, recommendations

The limitations of the research provide an interpretative framework for the results, which should be seen in this context. Their utilisation for policy purposes should also be based on this. The model change process of the Hungarian higher education (after several years of preparation) gained momentum in 2019 and was completed in administrative/normative terms in 2020 with the reorganisation of some universities into foundation operating forms and with the inclusion and declared assumption of the third mission and entrepreneurial university functions in the rules of operation. The literature shows that the first entrepreneurial universities date back centuries in the USA [97] and several decades in Europe [98]. The results of reform and transformation processes in higher education often take years or decades to become apparent. Therefore, the surprising results of the research in the light of the literature become explainable.

Several studies have confirmed that the university environment has a significant impact on entrepreneurial intention and entrepreneurial attitudes of students. This is supported by dozens of studies related to entrepreneurial education at universities, but other contextual elements have also been proven to influence these variables [26]. Even the research on testing the original EICM model came to this conclusion [23]. A significant proportion of these studies either treats the university environment and university support as a single variable or has collected data from higher education institutions where an entrepreneurial university operation and an entrepreneurial atmosphere have been present for years or decades. (In this respect, let us disregard research and studies that specifically examine *entrepreneurial education* functions, as they cover a much narrower field and thus the results obtained there will not be comparable to the present results.) Perhaps one of the most surprising results is that contrary to preliminary expectations (based on the literature), no significant correlation between entrepreneurial attitudes, entrepreneurial intentions and the university environment has been found. At the same time, there is a fairly strong positive relation between the variables of perceived behavioural control and university support. In addition, a covariance between the perceived university environment and the perceived university entrepreneurial support can also be detected. The above results can be explained in several possible ways. Among these explanations, the one with a theoretical background in organisational development seems to be the most convincing, considering all contextual circumstances, and this provides implications regarding future challenges for higher education policy makers.

The institutional characteristics influencing the factors that constitute the variables associated with each university ecosystem can be relatively easily aligned with the McKinsey 7S model [99], a framework for organizational effectiveness factors. The framework was developed primarily for operational efficiency analysis and forecasting at companies, but the method can also be applied to organisational analysis at other entities [100], such as government and higher education institutions. Based on the framework, the institutional factors influencing the university support variable can be classified as "*hard*" and the factors influencing the university environment variable as "*soft*". The soft elements are the shared values of the given institution, the set of skills and competences, the leadership and management style, and the motivations and attitudes of the staff. The hard elements include the institutional strategy, the organisational structure and the systems that influence day-to-day procedures and workflows. Perceived university environment factors (such as the atmosphere of the university, the inspiring effect of the university lecturers on starting a business, the inclusive events that foster entrepreneurial mindset, and the education for effective entrepreneurial mindset) are more likely to be influenced by the soft institutional elements. Based on experience, achieving the changes envisaged in these plans is only feasible in the longer term [100]. On the other hand, the hard elements defining the university support (available resources, university professional support and advice, and promoting market access for companies with a university background) can also deliver results in the shorter term. This requires management decisions, securing financial resources and finding the right external suppliers.

Based on the logic of the framework presented above, a possible explanation is outlined for the lack of a significant correlation between perceived university environment and entrepreneurial attitudes and entrepreneurial intention. Since the survey lacks a longitudinal dimension, the data collection can be considered a snapshot regarding the students. This snapshot is taken at a higher education institution that has undergone a model change process and is in transition. For this reason, it is likely that, on the one hand, reforms on the hard elements that can be implemented in the shorter term are already perceived as having an impact among students (positive and significant correlation between university support and perceived behavioural control). On the other hand, it is also likely that changes in the soft elements that can have an impact in the longer term have not yet reached the critical point where they can have a meaningful impact on the student community, which may explain the current lack of correlation between the university environment and student entrepreneurial attitudes and entrepreneurial intentions.

Since there is also covariance between the perceived university environment and the perceived university support based on the research results, the presented possible explanation also carries policy implications, which, accepting the above interpretation, can be summarised as follows:

- The transformation into an entrepreneurial university should be seen as a long-term process with regard to achieving the desired effects.

- Among the desired effects, those based on direct university support (capital, expertise, support for market entry) will be the earliest to be perceived, which will primarily enhance entrepreneurial intentions by influencing students' behavioural control.

- The development of a supportive university environment (inspiring atmosphere and teachers, effective curricular and extra-curricular teaching to foster entrepreneurial skills) is a time-consuming process and its results can only be measured in the long term.

- The level of development of the national and regional innovation system is also a key factor that needs to be taken into account in designing the elements and measures of the innovation ecosystem. In an emerging innovation country, different measures and entrepreneurships programs are needed than in a moderate or strong innovator country.

- At the initial stage in the process of transformation into an entrepreneurial university, the university support environment will not yet directly support students' entrepreneurial intentions, but international experience and research show that at some point in this transformation this will become perceivable and measurable.

- In the process of transformation into an entrepreneurial university, a key element of university governance is performance assessment, which is central to the competitive market environment of higher education. In terms of performance assessment, the present research suggests that it is worth considering a gradual introduction of the enhancement of student entrepreneurial attitude (as a performance indicator), since the conditions that promote this are developed gradually in the system.

- Due to the covariance between perceived university support and perceived university environment, we can identify interacting yet separate elements regarding students. It is evident that there is a greater scope and perspective for improvement in those factors which currently do little or nothing to foster student entrepreneurial intentions, although international experience shows that they are capable of doing so. At the same time, when university management perceives shortcomings in the university environment related to the entrepreneurial mindset, it will not be able to compensate with hard elements (more capital, more experts, more industrial partners). Conversely: the lack of capital and supporting market expertise cannot be compensated for by a supportive university environment, atmosphere, and entrepreneurial courses.

The essence of the 7S strategy developed by Tom Peters is summarised in six words by the author himself in the foreword to his book: *"Hard is soft. Soft is hard"* [99]. The results of the present research suggest that the above findings also prove valid for the transformation of higher education institutions into entrepreneurial universities. The development of an entrepreneur-friendly university environment based on soft elements is a time-consuming process, the results of which will be perceived in the student community only after the critical mass has been exceeded, and not immediately.

## Supporting information

**S1 File.**
(XLSX)

## Author Contributions

**Conceptualization:** Tibor Dőry.

**Data curation:** Attila Lajos Makai.

**Formal analysis:** Attila Lajos Makai.

**Methodology:** Attila Lajos Makai, Tibor Dőry.

**Project administration:** Attila Lajos Makai.

**Resources:** Attila Lajos Makai.

**Software:** Attila Lajos Makai.

**Supervision:** Tibor Dőry.

**Visualization:** Attila Lajos Makai.

**Writing – original draft:** Attila Lajos Makai.

**Writing – review & editing:** Tibor Dőry.